\documentclass{aipproc}
\usepackage{graphicx} 
\usepackage{amssymb}
\usepackage{amsmath}

\newcommand{\ba}{\begin{array}}
\newcommand{\ea}{\end{array}}
\newcommand{\be}{\begin{equation}}
\newcommand{\ee}{\end{equation}}
\layoutstyle{6x9}

\begin{document}

\title{Neutrino phenomenology and unparticle physics}
\author{\underline{J. Barranco}}{
address={Max-Planck-Institut f\"ur Gravitationsphysik 
(Albert-Einstein-Institut), Am M\"uhlenberg 1, D-14476 Golm, Germany
}
}
\author{A. Bola\~nos}{
address={Departamento de F\'{\i}sica, Centro de Investigaci{\'o}n y de
  Estudios Avanzados del IPN, Apartado Postal 14-740 07000
  M\'exico, D F, M\'exico}
}
\author{O. G. Miranda}{
address={Departamento de F\'{\i}sica, Centro de Investigaci{\'o}n y de
  Estudios Avanzados del IPN, Apartado Postal 14-740 07000
  M\'exico, D F, M\'exico}
}
\author{C. A. Moura}{
address={INFN Sezione di Napoli, Complesso Universitario Monte S. Angelo,
  Via Cintia, I-80126 Napoli, Italy}
}
\author{T. I. Rashba}{
address={Max-Planck-Institute for Solar System
Research, Katlenburg-Lindau, 37191, Germany and 
IZMIRAN, Troitsk, Moscow region, 142190, Russia}
}

\begin{abstract}
We show how neutrino data can be used in order to constrain the free parameters
of possible extensions to the standard model of elementary particles (SM). 
For definiteness, we focus in the recently proposed {\it unparticle} scenario. 
We show that neutrino data, in particular the MUNU experiment, 
can set stronger bounds than previous reported limits in 
the scale dimension parameter for certain region ($d > 1.5$). 
we compute the sensitivity of future neutrino experiments to 
unparticle physics such as future neutrino-electron scattering 
detectors, coherent neutrino-nuclei scattering as well as the ILC . 
In particular, we show that the measurement of coherent reactor 
neutrino scattering off nuclei provide a good sensitivity to 
the couplings of unparticle interaction with neutrinos and quarks.
Finally our results are compared with the current astrophysical limits.
\end{abstract}


\maketitle


\section{The $\nu-$SM}
The standard model of elementary particles (SM) is perhaps the most
successful model that has been experimentally tested. 
Nevertheless, the so called solar and atmospheric neutrino problems 
jeopardized the effectiveness of the SM. It took around 30 years of intense
debate and confrontation of theory and experiments to realize that
the correct mechanism needed to understand the deficit between 
expected neutrinos and the measured events
at terrestrial detectors is the neutrino oscillation mechanism 
\cite{Bilenky:2004xm}: 
neutrinos change their flavor due to the fact that
they are massive and their flavor states are a mixture of 
their mass states. 
This is the first evidence of physics beyond the SM and
consequently it requests to extend the SM in the leptonic sector.  
In summary, SM needs to incorporate: 
\begin{itemize}
\item Neutrinos masses $m_{\nu 1},m_{\nu 2},m_{\nu 3}$ and,
\item mixing in leptonic sector, that is, an equivalent to the CKM
matrix called $U_{PMNS}(\theta_{ij})$,
where the matrix $U_{PMS}$ is the mixing matrix between flavor states and 
mass states, i.e. $|\nu_\alpha \rangle = \sum_a U_{\alpha a}^* | \nu_a \rangle\,,a=1..3, \alpha=e,\,u,\tau$.
\end{itemize}
This will be refereed as the $\nu$-SM. The current values
for the masses and mixing angles can be found elsewhere \cite{Amsler:2008zzb}.
The objective of the present note, as part of the Proceeding of the 
CINVESTAV's Advanced Summer School, is to illustrate how neutrino 
phenomenology, within the $\nu-$SM, can constrain others possible 
extensions to the SM. As an example, we will focus on the recently
proposed {\it unparticle} scenario~\cite{Georgi:2007ek,Georgi:2007si}, where 
a scale invariant sector could exist above TeV energies that can couple
with the SM sector in the low-energy limit.
For the student's benefit, in this note we present a general overview of
how constrain nonstandard properties with neutrino.
For a more detailed discussion on the unparticle's parameter limits
by using the latest reactor neutrino data, we suggest to read \cite{Barranco:2009px}. 

\section{Constraining new physics with neutrino data}
It turns out that the biggest difficulty for detecting neutrino is
at the same time the biggest advantage: the neutrino interacts
only through weak interactions. Within the $\nu-$SM, the neutrino
doesn't have electric charge, or magnetic moment and any other
interaction will be automatically considered as nonstandard interaction.
That makes them the perfect candidate for detecting physics that is 
not considered within the SM because most of the times, in the low energy
limit, those extension will produce new charged or neutral currents, flavor
neutral changing currents (FCNC), or new neutrino properties.
This situation can be easily translated into an algorithm for constraining 
the parameters of a possible SM's extension: 

\begin{enumerate}
\item Star with $\nu$-SM: SM+neutrino masses and mixing. Compute standard processes. \label{uno} 
\item Propose a new theory: derive a new effective Lagrangian at low
energies, $\mathcal{L}^{NEW}(\epsilon, \eta,...)$, that introduces new
parameters besides those considered in the SM. \label{dos}
\item Compute physical processes: either scattering or decay processes 
($\frac{d\sigma^{New}}{dT}$,$\Gamma^{New}$).\label{tres}
\item Impose limits on the free parameters by comparing with experimental data
through an statistical analysis (usually using a $\chi^2$ analysis). \label{cuatro}
\end{enumerate}
For instance, the step number \ref{tres} can be directly be used
to compute the expected number of neutrinos.  
Neglecting for a moment the detector efficiency 
and resolution, it can be computed as
\begin{equation}
N_{\rm{events}}^{\rm{Theo}}(\eta,\epsilon...)=t\phi_0\frac{M_{\rm{detector}}}{M}
\int\limits_{E_{min}}^{E_{max}}dE_\nu
\int\limits_{T_{th}}^{T_{max}(E_\nu)}dT
\lambda(E_\nu)\frac{d\sigma}{dT}(E_\nu,T, \eta,\epsilon...)\,,
\label{Nevents}
\end{equation}
with $t$ the data taking time period, $\phi_0$ the total neutrino
flux, $M_{\rm{detector}}$ the total mass of the detector,
$\lambda(E_\nu)$ the normalized neutrino spectrum, $E_{max}$ the
maximum neutrino energy and $T_{th}$ the detector energy threshold.
The contribution due to new physics is encapsulated in the total cross section
\begin{equation}
\frac{d \sigma(E_\nu,T,\eta,\epsilon...)}{dT}=\frac{d \sigma^{SM}(E_\nu,T}{dT}
+\frac{d \sigma^{New}(E_\nu,T,\eta,\epsilon)}{dT}\,,
\end{equation}
which contains both, the SM and the new physics contribution obtained
from $\mathcal{L}^{NEW}(\eta,\epsilon...)$ including also the
interference terms, if they are not absent.
For the step number \ref{cuatro}, the comparison of the theoretical expected
events and the experimentally observed can be done by computing and
minimizing the $\chi^2$ function
\begin{equation}
\chi^2(\eta,\epsilon...)={\left(N_{\rm{events}}^{\rm{Exp}}-N_{\rm{events}}^{\rm{Theo}}(\eta,\epsilon...)\right)^2
\over \delta N_{\rm events}^2} \,,
\end{equation}
which automatically give us the allowed region for the free parameters $\{\eta,\epsilon...\}$ at some
confidence level.

\section{The case of unparticle physics}
Neutrino data can offer the possibility of studying unparticle
phenomenology in two ways: first one by studying the effects of
virtual unparticle exchanged between fermionic currents, second one by
studying the direct production of unparticles.  The neutrino-electron
and neutrino-nuclei scattering are examples where unparticle effects
of the first type are measurable, while single-photon production
($e^-e^+ \to \gamma\, X$) at LEP is an example of direct production of
unparticles.  Notice that, beside neutrinos ($\nu \bar \nu$), $X$ can
be any new hypothetical particle, in particular, unparticle stuff.  In
this case, neutrino production is the background reaction, because the
signatures for detection of unparticles are also the missing energy
and momentum.

Now let us perform our algorithm for the particular case of the {\it
unparticle} scenario.

\begin{itemize}
\item {\it Step 1}: Starting with the $\nu-$SM, the relevant processes are the
neutrino-electron scattering and the electron-positron annihilation to photon plus
missing energy. The respective $\nu$-SM cross sections are shown in table
\ref{tablaSM}. Still not observed coherent neutrino-nuclei scattering, $\nu+N\to\nu+N$, 
is shown also as a future perspective.
\begin{table}
\begin{tabular}{cc}
\hline
$\nu-$SM process & Cross section \\
\hline
$\nu_e+e\to \nu_e+e$ & ${d\sigma(E_{\nu}, T) \over dT}= {G_F^2 M_e \over 2\pi} ((g_V+g_A)^2+(g_v-g_A)^2(1-{T \over E_{\nu}})^2-(g_v^2-g_A^2) m_e {T \over E_{\nu}})$\\
$e^++e^-\to \nu\bar \nu +\gamma$& $\sigma^{\rm theo}_{\rm LEP}(s) =  \int {\rm d} x  \int {\rm d} c_{\gamma}
{2 \alpha \over \pi x s_\gamma }\left[
\left(1-{x \over 2}\right)^2+ {x^2 c_\gamma^2 \over 4} \right]\sigma_0^{\rm theo}(\hat{s})$\\
Coherent $\nu+N\to\nu+N$ &$\frac{d\sigma}{dT}=\frac{G_F^2 M}{2\pi}\left\{
(G_V+G_A)^2+\left(G_V-G_A\right)^2\left(1-\frac{T}{E_\nu}\right)^2-
\left(G_V^2-G_A^2\right)\frac{MT}{E_\nu^2}\right\}$ \\
\hline
\end{tabular}
\caption{Cross section for the relevant SM processes. Here $G_F$ is the Fermi constant,
$g_V, g_A, g_v^{n,p}, g_A^{n,p}$ are the SM coupling constants to electron, protons and neutrons 
respectively,$M_e$ the electron mass, $M$ the nucleon mass, 
$G_V=\left[g_V^p Z+g_V^n N\right]F_{nucl}^V(Q^2)$,
$G_A=\left[g_A^p\left(Z_+-Z_-\right)+
g_A^n\left(N_+-N_-\right)\right]F_{nucl}^A(Q^2)$. $F_{nucl}^V(Q^2)$ is the nuclear form factor,
$Z$ the atomic number, and $N$ the number of neutron in the nuclei. $\pm$ in $N,Z$ refers to 
the polarization.}\label{tablaSM}
\end{table}
\item {\it Step 2:} Take an extension to the SM, in this case we will focus in the unparticle
scenario. From the phenomenological point of view, the relevant
point is that at the low energy regime, there will be an effective interactions for 
the scalar and vector unparticle operators with the SM fermion fields which are expressed in the effective 
Lagrangian:
\begin{eqnarray}\label{scalar}
{\cal L}_{{\cal U}_S}(\lambda_{0f},d, \Lambda) &=&
\lambda_{0f}\frac{1}{\Lambda^{d-1}}\,\bar f f \,{\cal O_U} +
\lambda_{0\nu}^{\alpha\beta}\frac{1}{\Lambda^{d-1}}\,\bar \nu_\alpha
\nu_\beta \,{\cal O_U}\\
{\cal L}_{{\cal U}_V}(\lambda_{1f},d, \Lambda) &=&
\lambda_{1f}\frac{1}{\Lambda^{d-1}}\,\bar f \gamma_\mu f \,{\cal O}_{\cal U}^\mu +
\lambda_{1\nu}^{\alpha\beta}\frac{1}{\Lambda^{d-1}}\,\bar \nu_\alpha
\gamma_\mu \nu_\beta \,{\cal O}_{\cal U}^\mu \,,
\end{eqnarray}
for the scalar and vectorial interaction. Here $\alpha=e,\mu,\tau$ and $f=e,u,d$. 
The parameters of the model are therefore $(\lambda_{if},d, \Lambda)$. 
From now and on, we will fix the energy scale where the theory is invariant to $\Lambda=1$~TeV. 
\item {\it Step 3}: Compute the physical processes. With the help of the Lagrangian 
eq. (\ref{scalar}), we compute the neutrino-electron scattering mediated 
by the scalar unparticle  and the vector unparticle. For the case of the vector interaction, 
there is an interference term between the SM model cross section and the unparticle sector.
All relevant expressions for $\nu+e$ scattering are collected in table \ref{scat}.
The $\nu_e+e$ scattering is an example of the role of unparticle as intermediate of a new interaction.
But the unparticle stuff can be produced directly at accelerators through for instance $e^++e^-\to \mathcal{U} + \gamma$.
In this case, the relevant cross section is computed. (See bottom line of Table \ref{scat}). 

\begin{table}
\begin{tabular}{cc}
\hline
Unparticle interaction& Cross section \\
\hline
Scalar &$\frac{d\sigma_{{\cal U}_S}}{dT}=\frac{[g^{\alpha\beta}_{0e}(d)]^2}{\Lambda^{(4d-4)}}
\frac{2^{(2d-6)}}{\pi E_\nu^2}(m_eT)^{(2d-3)}(T+2m_e)$\\
Vectorial& 
$\frac{d\sigma_{{\cal U}_V}}{dT}=\frac{1}{\pi}
\frac{[g_{1e}^{\alpha\beta}(d)]^2}{\Lambda^{(4d-4)}}
2^{(2d-5)}(m_e)^{(2d-3)}(T)^{(2d-4)}
\left[1 + \left(1-\frac{T}{E_\nu}\right)^2-\frac{m_eT}{E_\nu^2}\right]$\\
Interference vectorial & 
$\frac{d\sigma_{{\cal U}_V-SM}}{dT}=\frac{\sqrt{2} G_F}{\pi} 
\frac{g_{1e}(d)}{\Lambda^{(2d-2)}}\, (2m_e T)^{(d-2)}
m_e \left\{
g_L + g_R \left(1-\frac{T}{E_\nu}\right)^2 - \frac{\left(g_L+g_R\right)}{2} 
\frac{m_e T}{E_\nu^2}\right\}$\\
\hline
$e^++e^-\to \mathcal{U} + \gamma$ &$\frac{d\sigma_{\gamma{\cal U}}}{dx}
=\int_{y_{min}}^{y_{max}}\frac{A_d}{(4\pi)^2}
\left(\frac{\lambda_{1e}e}{\Lambda}\right)^2
\left[\frac{s(1-x)}{\Lambda^2}\right]^{(d-2)}\frac{x^2+x^2y^2+4(1-x)}{x(1-y^2)}dy$\\

\hline
\end{tabular}
\caption{Cross section mediated by scalar and vectorial unparticle propagators
for the $\nu_e+e$ scattering. We have defined 
$g^{\alpha\beta}_{if}(d)=\frac{\lambda_{i\nu}^{\alpha\beta}\lambda_{if}}{2\sin(d\pi)}\frac{16\pi^{5/2}}{(2\pi)^{2d}}
\frac{\Gamma(d+1/2)}{\Gamma(d-1)\Gamma(2d)}$,$ i=0,1$ for scalar and vectorial interaction respectively.
Cross section for the single-photon production in 
electron-positron collisions. Here $A_d = \frac{16\pi^{5/2}}{(2\pi)^{2d}}
\frac{\Gamma(d+1/2)}{\Gamma(d-1)\Gamma(2d)}$.}\label{scat}
\end{table}

Another interesting experimental proposal is the coherent neutrino nuclei interaction \cite{Barranco:2005yy}. 
In this case, if $qR < 1$ ($R$ nuclei radii, $q$ the transfered momentum), the  $\nu$ "sees'' 
the nuclei as a point and scatters coherently on it as a whole. This effect has not been measured yet, 
but there are some experimental proposal planned to detect this process \cite{Wong:2005vg,coherentproposal,Barbeau:2007qi}. 
The effective Lagrangian \ref{scalar} would give also contributions to this process and therefore we have 
considered it too in our analysis. It is shown that if such process is detected it would place
strong constraints on unparticle interaction. 
The corresponding cross sections are collected in Table \ref{coherent}. Note
that there is again an interference term between the SM sector and the unparticle sector for
the case of a vector intermediate unparticle.

\begin{table}
\begin{tabular}{cc}
\hline
Type&Cross section \\
\hline
Scalar&$\frac{d\sigma_{{\cal U}_S}^{\nu N}}{dT}=\frac{1}{\Lambda^{(4d-4)}} \,
\frac{2^{(2d-6)}}{\pi E_\nu^2} 
\left[g_{0u}(d)(2Z+N)+g_{0d}(d)(Z+2N)\right]^2
(m_A T)^{(2d-3)}(T+2m_A)$\\
 
Vectorial&$  \frac{d\sigma_{{\cal U}_V}^{\nu N}}{dT}=\frac{2^{(2d-5)}}{\pi\Lambda^{(4d-4)}}
  \frac{m_A}{(m_AT)^{(2d-4)}}
  \left[g_{1u}(d)(2Z+N)+g_{1d}(d)(Z+2N)\right]^2\left[1  + \left(1-\frac{T}{E_\nu}\right)^2-\frac{m_AT}{E_\nu^2}\right]$\\

SM+U&$\frac{d\sigma_{{\cal U}_V-SM}^{\nu N}}{dT}=
\frac{\sqrt{2} 2^{d-1} G_F}{\pi} \frac{\left[g_{1u}(d)(2Z+N)+g_{1d}(d)(Z+2N)\right]}
{\Lambda^{(2d-2)}}\frac{m_A\left(g_V^p Z+g_V^n N\right)}{(m_AT)^{(d-2)}}
\left[ 1 + \left(1-\frac{T}{E_\nu}\right)^2-\frac{m_AT}{E_\nu^2}\right]$\\
\hline
\end{tabular}
\caption{Cross section mediated by scalar and vectorial unparticle propagators
for the coherent $\nu_e+N$ scattering.}\label{coherent}
\end{table}
Now it is possible to compute the expected number of events as a function of the 
free parameters $N_{events}^{Theo}(\lambda_{if},d)$ with the help of expressions collected in 
Tables \ref{scat}-\ref{coherent}.
\item {\it Step 4}: Now let us compute our $\chi^2(\lambda_{if},d)$ where the experimental 
number of events for the processes collected in table \ref{scat} are well known (summarized in \cite{Barranco:2007ej}). 
For the case of coherent neutrino-nuclei scattering (Table \ref{coherent}) we will take
for definiteness the TEXONO detector \cite{Wong:2005vg} 
neglecting for the moment the resolution function.
One more time, for details of the full analysis we refer to \cite{Barranco:2009px}.
\end{itemize}

\begin{figure}
        \includegraphics[width=0.3\columnwidth]{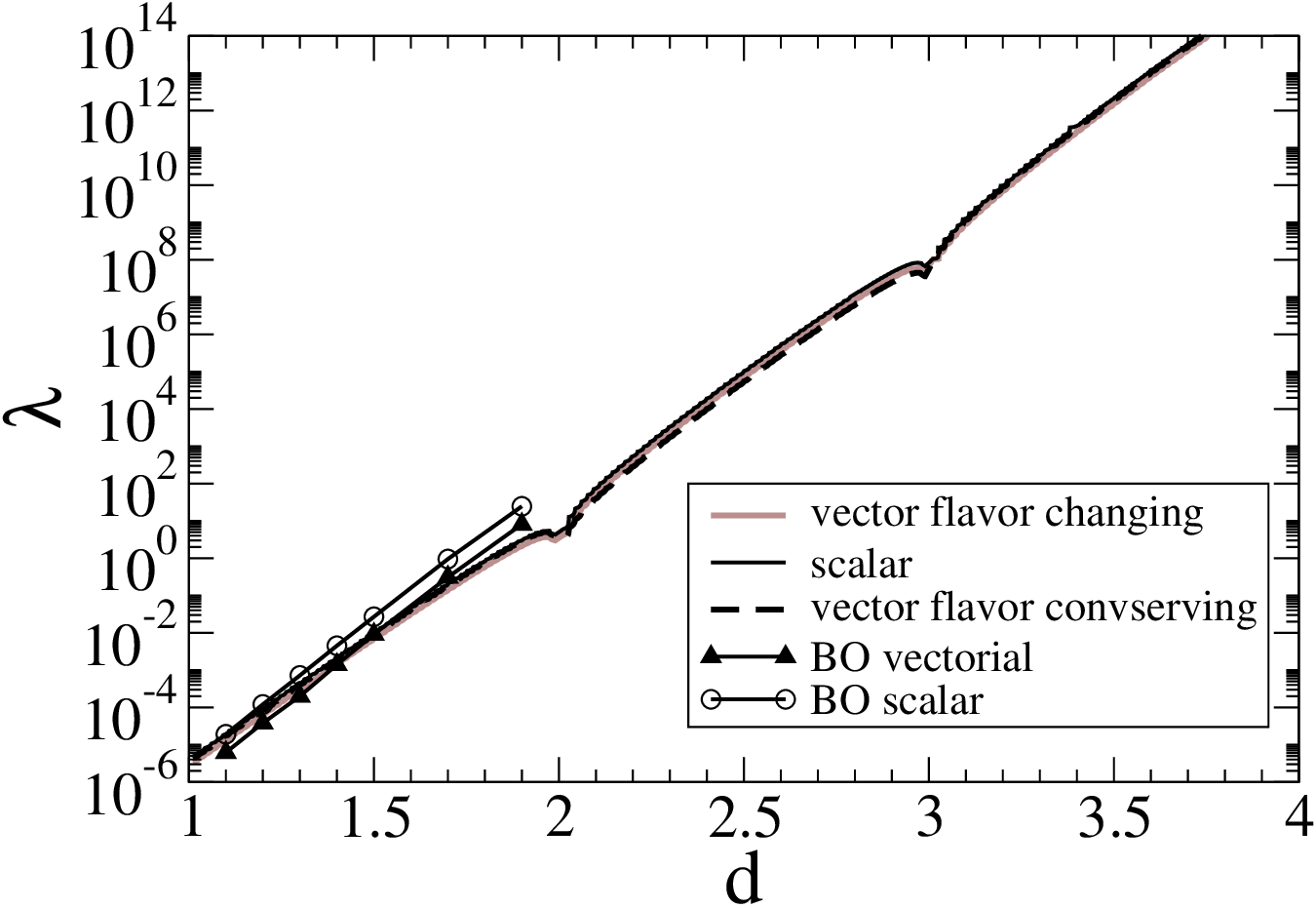}
        \includegraphics[width=0.3\columnwidth]{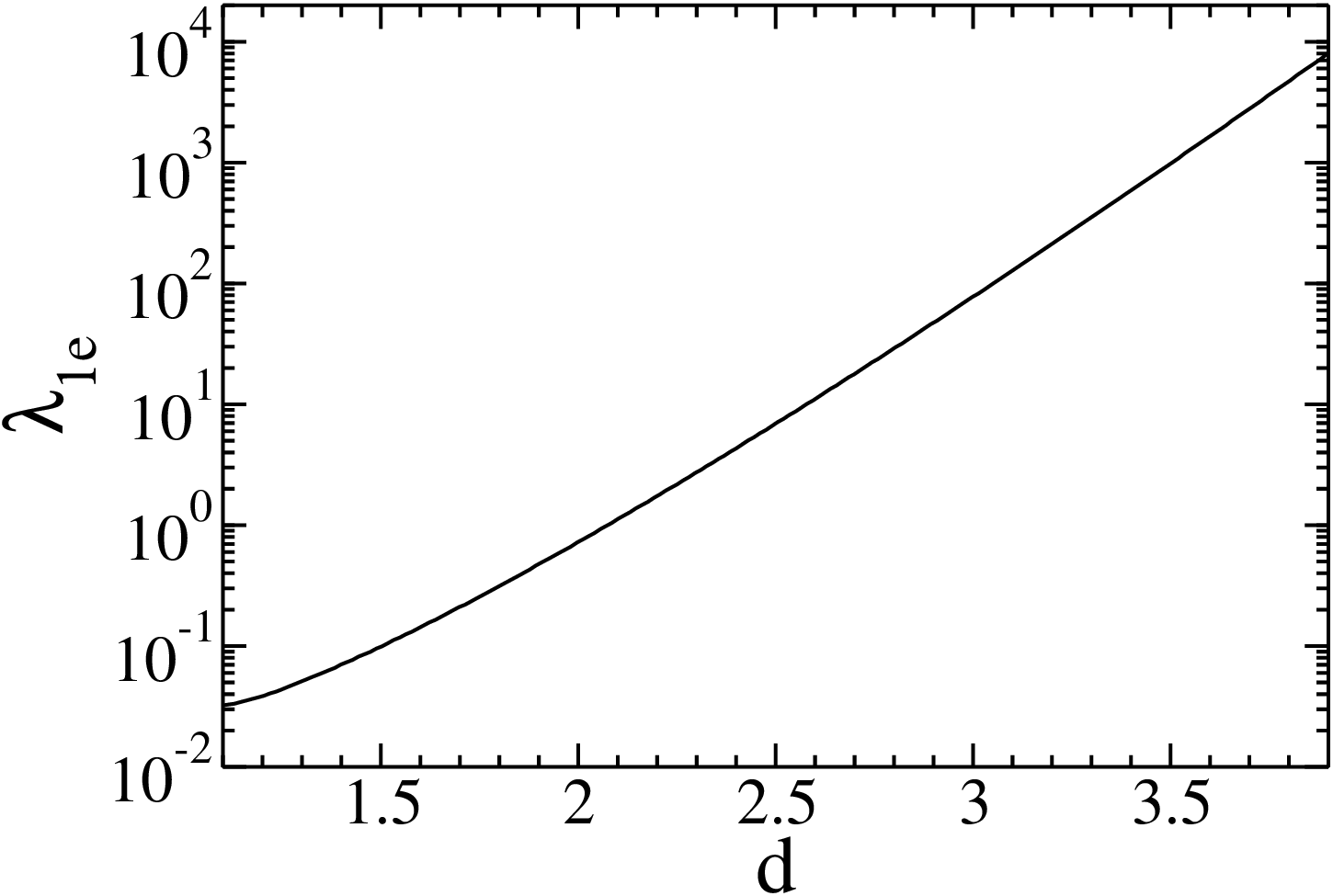}
        \includegraphics[width=0.3\columnwidth]{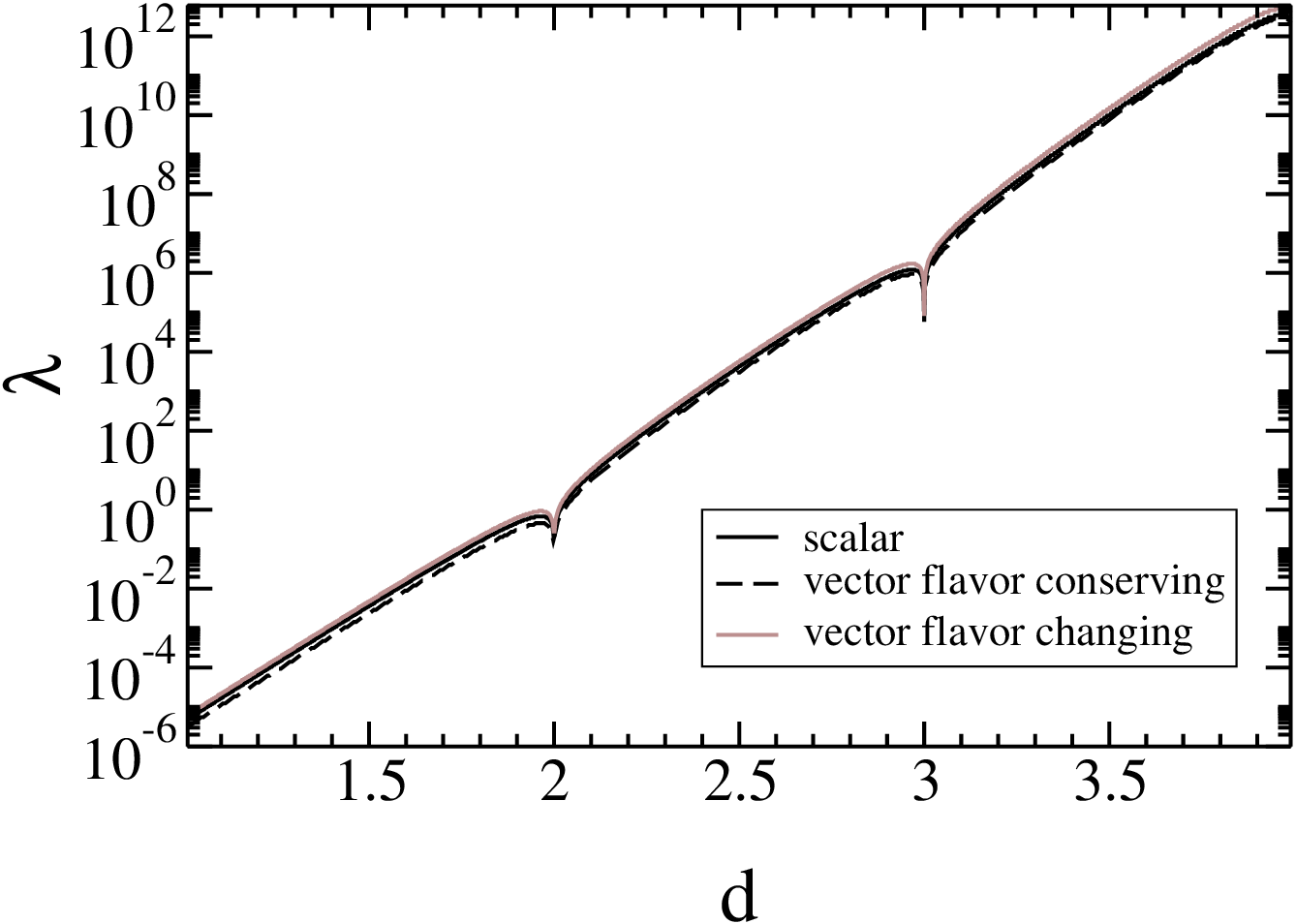}
\caption{Left: Limits on the parameters $d$ and
  $\lambda_{0,1}=\sqrt{\lambda^{e\beta}_{0,1\nu}\lambda_{0,1e}}$ (90
  \% CL) from the MUNU experiment for the scalar unparticle case
  (black solid line) and for the vector unparticle cases, both for
  flavor changing currents (grey solid line) and for the flavor
  conserving conserving case (dashed line). Previous bounds obtained
  by Balantekin and Ozansoy (BO)~\cite{Balantekin:2007eg} (dots and
  triangles) are shown for comparison. The present analysis based
  on the MUNU data gives stronger constraints on $\lambda_{0,1}$ for
  values of $d>1.5$. Middle: Limits obtained with LEP data. Right: Sensitivity
for unparticle in the coherent $\nu +N \to \nu+N$ scatering TEXONO's proposal.}
\label{fig:nu-e}
\end{figure}

{\bf Results:}
We have obtained the constrains illustrated in Fig. \ref{fig:nu-e} at 90\% C.L. for the scale dimension $d$ and the
coupling $\lambda_{0,1}=\sqrt{\lambda^{e\beta}_{0,1\nu}\lambda_{0,1e}}$. 
The values below the lines are allowed. The first two plots show the limits
imposed  by the current MUNU and LEP data. The third plot 
shows the sensitivity that an experiment able to detect the coherent neutrino-nuclei scattering.

\begin{table}[!t]  
        \begin{tabular}{c|c|c|c|c}
            \hline   
            d  & $\nu-e$ scattering & E\"otv\"os \cite{Deshpande:2007mf} & Long range \cite{GonzalezGarcia:2008wk} & SN1987A \cite{Hannestad:2007ys} \\
            \hline
            \hline 
            1.1 &$2.0\times 10^{-5}$&$6.3\times 10^{-19}$&$2.8\times 10^{-23}$
&$9.1\times 10^{-11}$\\
            1.5 &$9.7\times 10^{-3}$&$1.7\times 10^{-12}$&$5.7\times 10^{-12}$&
$5.7\times 10^{-9}$\\
            2.1 &$40.$&$1.1\times 10^{-2}$&$6.0\times 10^{5}$&
$2.9\times 10^{-6}$\\
            2.5 &$5.5\times 10^{4}$&$4.8\times 10^{4}$&$1.1\times 10^{17}$&
$1.8\times 10^{-4}$\\
            3.1 &$1.2\times 10^{9}$&$3.3\times 10^{14}$&$1.1\times 10^{34}$&
$9.9\times 10^{-2}$\\
            3.5 &$2.1\times 10^{12}$&$1.5\times 10^{21}$&$3.2\times 10^{45}$&
$6.1$\\
            3.9 &$1.1\times 10^{15}$&$6.2\times 10^{27}$&$5.8\times 10^{56}$&
$414.3$\\
            \hline
        \end{tabular}  
    \caption{Constraints on the vector coupling $\lambda_1$ from the
 neutrino electron scattering experiment, and from astrophysical
 limits.}  
    \label{tablavectorial}
\end{table}
\subsection{Astrophysical neutrino and unparticle}
There is another interesting issue related with neutrinos. They are copiously produced 
in most of the astrophysical objects: the Sun, neutrons star, supernova explosions, merger of 
Neutron stars, AGN etc, all those astrophysical phenomena involve neutrino production. 
A very small change in the neutrino properties or interactions will enhance or decrease the
number of neutrinos produced due to the large amount of matter and high energies involved. 
This fact has been used successfully in order to constrain neutrino properties and/or new interactions.
The unparticle case was not the exception and currently there are several limits
obtained. In table \ref{tablavectorial} we show some limits and we include the limits
obtained from our analysis. Despite some of the limits are orders of magnitude stronger than
the ones obtained by terrestrial experiments, there are assumptions made and, in that
respect, a direct measurement always offers a clean determination of free parameters.
For a more extensive discussion of the assumptions made see \cite{Barranco:2009px}.

In summary, we have illustrated how neutrino data can offer a clean way for constraining 
free parameters of new extensions to the standard model of elementary particles. We used
as a test the {\it unparticle} scenario and we see that present and future experiments in 
neutrino physics are competitive and complementary to other high energy and astrophysics 
observations.

{\bf Acknowledges} JB thanks the organizing committee of the EAV for the
opportunity of giving this talk. This work has been supported by CONACyT,
SNI-Mexico and PAPIIT Grant No. IN104208.


\begin{thebibliography}{99}
\bibitem{Bilenky:2004xm}
  S.~M.~Bilenky,
  Phys.\ Scripta {\bf T121}, 17 (2005).
\bibitem{Amsler:2008zzb}
  C.~Amsler {\it et al.}  [Particle Data Group],
  Phys.\ Lett.\  B {\bf 667}, 1 (2008).

\bibitem{Georgi:2007ek}
  H.~Georgi,
  Phys.\ Rev.\ Lett.\  {\bf 98}, 221601 (2007).

\bibitem{Georgi:2007si}
  H.~Georgi,
  Phys.\ Lett.\  B {\bf 650}, 275 (2007).

\bibitem{Barranco:2009px}
  J.~Barranco, A.~Bolanos, O.~G.~Miranda, C.~A.~Moura and T.~I.~Rashba,
  Phys.\ Rev.\  D {\bf 79}, 073011 (2009)

\bibitem{Barranco:2005yy}
  J.~Barranco, O.~G.~Miranda and T.~I.~Rashba,
  JHEP {\bf 0512}, 021 (2005).



\bibitem{Wong:2005vg}
  H.~T.~Wong, H.~B.~Li, J.~Li, Q.~Yue and Z.~Y.~Zhou,
  J.\ Phys.\ Conf.\ Ser.\  {\bf 39} (2006) 266.


\bibitem{coherentproposal}
  K.~Scholberg,
  Phys.\ Rev.\ D {\bf 73} (2006) 033005.
   J.~Serreau and C.~Volpe,
   Phys.\ Rev.\ C {\bf 70} (2004) 055502.
  A.~Bueno, M.~C.~Carmona, J.~Lozano and S.~Navas,
  Phys.\ Rev.\ D {\bf 74} (2006) 033010.

\bibitem{Barbeau:2007qi}
  P.~S.~Barbeau, J.~I.~Collar and O.~Tench,
  JCAP {\bf 0709}, 009 (2007).


\bibitem{Barranco:2007ej}
  J.~Barranco, O.~G.~Miranda, C.~A.~Moura and J.~W.~F.~Valle,
  Phys.\ Rev.\  D {\bf 77}, 093014 (2008).

\bibitem{Balantekin:2007eg}
  A.~B.~Balantekin and K.~O.~Ozansoy,
  Phys.\ Rev.\  D {\bf 76}, 095014 (2007).
\bibitem{Deshpande:2007mf}
  N.~G.~Deshpande, S.~D.~H.~Hsu and J.~Jiang,
  Phys.\ Lett.\  B {\bf 659}, 888 (2008).


\bibitem{GonzalezGarcia:2008wk}
  M.~C.~Gonzalez-Garcia, P.~C.~de Holanda and R.~Zukanovich Funchal,
  JCAP {\bf 0806}, 019 (2008).

\bibitem{Hannestad:2007ys}
  S.~Hannestad, G.~Raffelt and Y.~Y.~Y.~Wong,
  Phys.\ Rev.\  D {\bf 76}, 121701(R) (2007).

\end{thebibliography}
\end{document}